\documentclass{appolb}

\newcommand{\be}{\begin{equation}}
\newcommand{\ee}{\end{equation}}

\usepackage{graphicx}
\begin{document}
\title{The non-ordinary Regge behavior of the $f_0(500)$ meson
\thanks{Presented at ``Excited QCD''. Bjeliasnica Mountain, Sarajevo, 2-8 February 2014.}%
% you can use '\\' to break linesls
}
\author{J. R. Pel\'aez\footnote{speaker}, 
\vspace{-.4cm}
\address{Dpto. de F\'isica Te\'orica II, Universidad Complutense de Madrid, 28040, Spain}
\\  \vspace{.2cm}
J.~T.~Londergan
\vspace{-.4cm}
\address{CEEM, Indiana University, Bloomington, IN 47403, USA\\
Physics Department  Indiana University, Bloomington, IN 47405, USA}
\\  \vspace{.2cm}
J.~Nebreda
\vspace{-.4cm}
\address{Yukawa Institute for Theoretical Physics, Kyoto University, Kyoto 606-8502,
Japan}
\\ \vspace{.2cm}
A.~P.~Szczepaniak
\vspace{-.4cm}
\address{CEEM, Indiana University, Bloomington, IN 47403, USA\\
Physics Department  Indiana University, Bloomington, IN 47405, USA}
\vspace{-.3cm}
}
%{rtjh
%\address{affiliation}
%}
\maketitle
\begin{abstract}
%\vspace{-.9cm}
We review how the Regge trajectory of an elastic resonance can be obtained just from its pole position and coupling, by means of a dispersive formalism. This allows to deal correctly with the finite widths of resonances in Regge trajectories. For the $\rho(770)$ meson this method leads to the ordinary linear Regge trajectory with a universal slope. In contrast, for the $f_0(500)$ meson, the resulting Regge trajectory is non-linear and with much smaller slope. This is another strong indocation of the non-ordinary nature of the lightest scalar meson.
\vspace{-.4cm}
\end{abstract}
%\PACS{PACS numbers come here}
  
\vspace{-.2cm}

\section{Introduction}\label{aba:sec1}

In a recent work \cite{Londergan:2013dza}, we used the analytic properties of amplitudes in the complex angular momentum plane to study the Regge trajectory associated to an elastic resonance. In principle, the form of these trajectories can be used to discriminate between the underlying QCD mechanisms that generate these resonances. Actually, linear $(J,M^2)$
trajectories relating the angular momentum $J$ and the mass squared are intuitively interpreted in terms of quark-antiquark states, since they are easily obtained from of the rotation 
of a flux tube connecting a quark and an antiquark. 
Strong deviations from this linear behavior would suggest a rather different nature and the scale of
the trajectory would also indicate the scale of the mechanism responsible for the presence of a resonance. 

In particular,  we have studied in \cite{Londergan:2013dza} 
the trajectories of the lightest resonances in 
elastic $\pi\pi$ scattering: the $\rho(770)$, which is a well established
ordinary $\bar qq$ state, and the $f_0(500)$ or $\sigma$ meson, whose nature 
is still under debate and 
whose resulting trajectory, as we will see, 
 does not follow the ordinary linear  $(J,M^2)$ trajectories.
Actually, scalar mesons, and particularly the sigma are not included 
in those linear fits \cite{Anisovich:2000kxa}, or its huge width is used as the uncertainty in the mass, so that it could be accommodated easily.  But as we will see, the width is part of the Regge trajectory, and considering the width as a mass uncertainty is not really justified.

\section{Regge trajectory from a resonance pole}

An elastic $\pi\pi$ partial wave near a Regge pole reads
\be
t_l(s)  = \beta(s)/(l-\alpha(s)) + f(l,s),
\label{Reggeliket}
\ee
where $f(l,s)$ is a regular function of $l$, and the Regge trajectory $\alpha(s)$ and 
residue $\beta(s)$ are analytic functions, the former having a cut along the real axis for $s > 4m_\pi^2$.

The analytic properties of $\alpha(s)$ and $\beta(s)$ together with the elastic unitarity condition imply that they are constrained by the follwing system of coupled dispersion relations~\cite{Chu:1969ga}:
\begin{eqnarray}
\mbox{Re}\, \alpha(s) & =&   \alpha_0 + \alpha' s +  \frac{s}{\pi} PV \int_{4m_\pi^2}^\infty ds' \frac{ \mbox{Im}\,\alpha(s')}{s' (s' -s)}, \label{iteration1}\\
\mbox{Im}\,\alpha(s)&=&  \frac{ \rho(s)  b_0 \hat s^{\alpha_0 + \alpha' s} }{|\Gamma(\alpha(s) + \frac{3}{2})|}
 \exp\Bigg( - \alpha' s[1-\log(\alpha' s_0)]\nonumber\\
&&+ \frac{s}{\pi} PV\!\int_{4m_\pi^2}^\infty\!\!ds' \frac{ \mbox{Im}\,\alpha(s') \log\frac{\hat s}{\hat s'} + \mbox{arg }\Gamma\left(\alpha(s')+\frac{3}{2}\right)}{s' (s' - s)} \Bigg), 
\label{iteration2}
 \end{eqnarray}
where $PV$ denotes ``principal value'' and $\alpha_0, \alpha'$ and $b_0$ are free parameters 
to be determined phenomenologically. For the $f_0(500)$-meson, one can also make explicit in $\beta(s)$ the Adler-zero required by chiral symmetry. In that case $b_0$ will not be dimensionless.

\section{$\rho(770)$ and $f_0(500)$ trajectories}
  
\begin{figure}
\hspace{-7mm}
\includegraphics[scale=0.71,angle=-90]{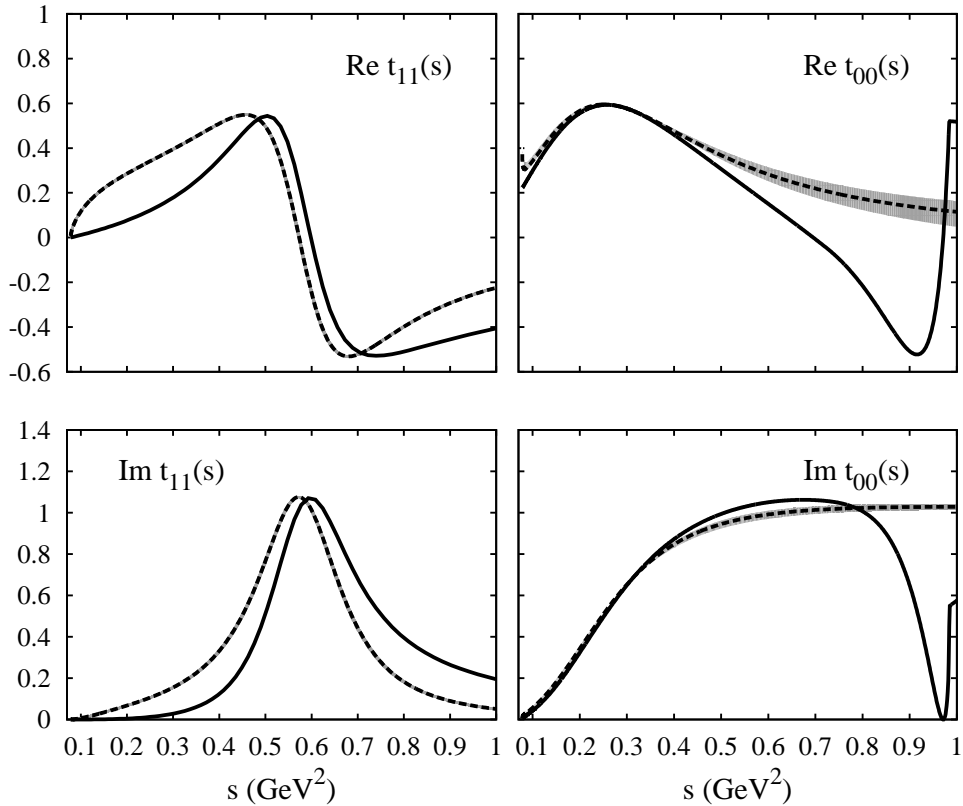}
 \caption{\rm \label{fig:ampl} 
  Partial waves $t_{lI}$ with $l=1$ (left panels) and $l=0$ (right panels). 
  Solid lines represent the amplitudes from the dispersive analisis of data in \protect{\cite{GarciaMartin:2011jx}}. Their corresponding resonance poles 
  are the only input for the constrained Regge-pole amplitudes shown with dashed curves. The gray bands cover only the uncertainties due to the uncertainties in the inputs. }
\end{figure}

For a given set of $\alpha_0, \alpha'$ and $b_0$ parameters we solve the system of Eqs.~(\ref{iteration1}) and (\ref{iteration2}) iteratively. The value of the parameters is fixed by fitting only three inputs, namely, the real and imaginary parts of the resonance pole position $s_M\simeq(M_R-i \Gamma/2)^2$, where $M_R$ and $\Gamma_R$ are the pole mass and width of the resonance, together with the absolute value of the pole residue $|g_M|$. Namely, we fit the resonance pole on the second Riemann sheet to:
$\beta_M(s)/(l  - \alpha_M(s))\rightarrow |g^2_M|/(s-s_M)$, 
with $l=0,1$ for $M=\sigma,\rho$. The pole parameters are taken from a precise dispersive representation of $\pi\pi$ scattering data \cite{GarciaMartin:2011jx}. 
In Fig.~\ref{fig:ampl} we compare, on the real axis, the partial waves obtained from data in \cite{GarciaMartin:2011jx} with the Regge amplitudes on the real axis obtained just from the pole fit. They do not need to overlap since they are only constrained to agree at the resonance pole. Nevertheless, we find a fair agreement in the resonant region. As expected, it deteriorates as we approach threshold or the inelastic region, particularly in the case of the $S$-wave due to the interference with the $f_0(980)$. But note that in the resonance region the Regge amplitude describes the empirical curve even better in the scalar than in the vector case.

\begin{figure}
\vspace{-3mm}
\hspace{-5mm}\includegraphics[scale=0.6,angle=-90]{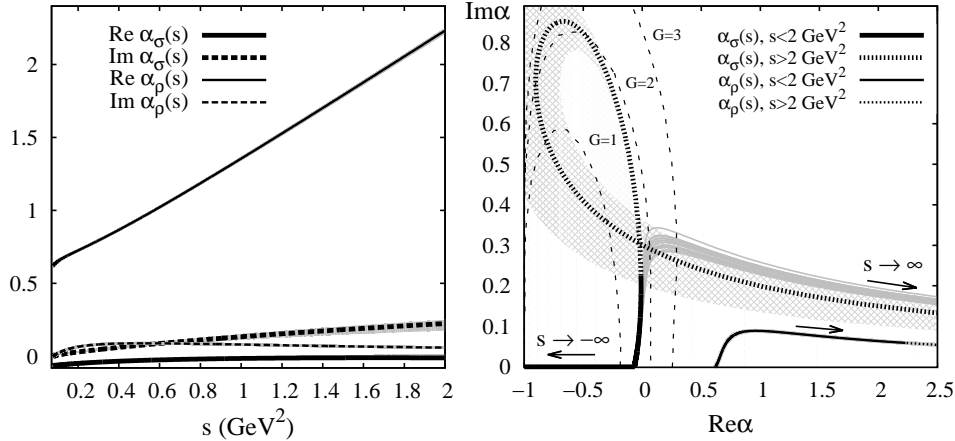}
 \caption{\rm \label{fig:trajectories} 
  (Left) $\alpha_\rho(s)$ and $\alpha_\sigma(s)$ Regge trajectories, 
from our constrained Regge-pole amplitudes. 
 (Right) $\alpha_\sigma(s)$ and $\alpha_\rho(s)$
in the complex plane. 
At low and intermediate energies (thick continuous lines), the trajectory of the $\sigma$ is similar to those of Yukawa potentials $V(r)=-{\rm G} a \exp(-r/a) /r$ \cite{Lovelace} (thin dashed lines). Beyond 2 Gev$^2$
we plot our results as thick discontinuous lines because they should be considered just as extrapolations.  }
\end{figure}

Therefore, once we have checked that fitting just the pole with a Regge like amplitude still
provides a fairly good description of the amplitude, we are ready to show their associated
Regge trajectories calculated with our dispersive formalism. In particular, in the left panel of Fig.~\ref{fig:trajectories} the resulting Regge trajectories, whose parameters are given in Table \ref{aba:tbl2}. The imaginary part of $\alpha_\rho(s)$ is much smaller than the real part, and the latter grows linearly with $s$. Taking into account our approximations, and that our error bands only reflect
the uncertainty in the input pole parameters, the agreement with previous determinations is remarkable: $\alpha_\rho(0)=0.52\pm0.02$~\cite{Pelaez:2003ky}, $\alpha_\rho(0)=0.450\pm0.005$ 
\cite{PDG}, $\alpha'_\rho\simeq 0.83\,$GeV$^{-2}$ \cite{Anisovich:2000kxa}, $\alpha'_\rho=0.9\,$GeV$^{-2}$ \cite{Pelaez:2003ky}, or $\alpha'_\rho\simeq 0.87\pm0.06$GeV$^{-2}$, \cite{Masjuan:2012gc}.

\begin{table}
\vspace{-3mm}
\begin{center}
\begin{tabular}{|c|ccc|}\hline
& $\alpha_0$ & $\alpha'$ (GeV$^{-2}$)  & \hspace{5mm}$b_0$\hspace{5mm} \\\hline%\colrule
$\rho(770)$ & $0.520\pm0.002$  & $0.902\pm0.004$ & $0.52$ \\
$f_0(500)$ &  $-0.090\,^{+\,0.004}_{-\,0.012}$ & $0.002^{+0.050}_{-0.001}$ & $0.12$ GeV$^{-2}$\\
\hline
\end{tabular}
\caption{ Parameters of the $\rho(770)$ and $f_0(500)$ trajectories}
\label{aba:tbl2}
\end{center}
\vspace{-4mm}
\end{table}
In contrast, the $f_0(500)$ trajectory is not evidently linear and its slope is about two orders of magnitude smaller than that of the linear trajectories of ordinary quark-antiquark  resonances, {\it e.g.}, $\rho$, $a_2$, $f_2$ or $\pi_2$. 
 This provides strong support for a non-ordinary nature of the $\sigma$ meson. Moreover, the resulting scale of tens of MeV or at most hundreds, for the slope, is more typical of meson physics than of quark-antiquark interactions.  In addition, the  
 tiny slope excludes the possibility 
 that any of the known isoscalar resonance may lie on its  trajectory.  To test the robustness of this observation we have checked that our results are very stable 
within the uncertainties of the pole parameters that we used as input. In addition we have
tried to impose a typical size linear trajectory on the $\sigma$, but that deteriorates
the fit to the $\sigma$ pole and particularly to the coupling,
so the resulting amplitude in the physical region is qualitatively very different from the observations. 

Note that with our formalism, we are dealing correctly with the huge $f_0(500)$ width, and, by no means should be considered as an uncertainty in the mass.

Furthermore, in Fig.~\ref{fig:trajectories} we show the striking similarities
 between the $f_0(500)$ trajectory and those of Yukawa potentials
in non-relativistic scattering \cite{Lovelace}. From the Yukawa G=2 curve in that plot, which lies closest to our result for the $f_0(500)$, we can estimate $a\simeq 0.5 \,$GeV$^{-1}$, following \protect \cite{Lovelace}. This could be compared, for instance, to the S-wave $\pi\pi$ scattering length $\simeq 1.6\, $GeV$^{-1}$. Thus it seems that the range of a Yukawa potential that would mimic our low energy results is comparable but smaller than the $\pi\pi$ scattering length in the scalar isoscalar channel. Of course, our results are most reliable at low energies (thick continuous line) and the extrapolation should be interpreted cautiously. Nevertheless, our results suggest that the $f_0(500)$ looks more like a low-energy resonance of a short range potential,  {\it e.g.}\ between pions, than a bound state of a confining force between a quark and an antiquark. 

In summary, our formalism and the results for the $f_0(500)$  explains why the lightest scalar meson has to be excluded from the ordinary linear Regge fits of ordinary mesos.
\section{Acknowledgements} 
JRP wants to thank the organizers of Excited2014 for creating such a stimulating workshop.
JRP and JN are supported by the Spanish project FPA2011-27853-C02-02. JN acknowledges funding by the Deutscher Akademischer Austauschdienst (DAAD), the Fundaci\'on Ram\'on Areces and the hospitality of Bonn and Indiana Universities. APS\ is supported in part by the U.S.\ Department of Energy under Grant DE-FG0287ER40365. JTL is supported by the U.S. National Science Foundation under grant 
PHY-1205019.

\end{document}